\documentclass{acm_proc_article-sp}
\makeatletter
\def\@copyrightspace{\relax}
\makeatother

\usepackage{bm}
\newcommand{\tabincell}[2]{\begin{tabular}{@{}#1@{}}#2\end{tabular}}
\usepackage{array}
\usepackage{graphicx}
\usepackage{tikz}

\newlength{\onesixth}
\newcolumntype{C}{>{\begin{minipage}{2\onesixth}\begin{center}}{c}<{\end{center}\end{minipage}}}
\newcolumntype{D}{>{\begin{minipage}{3\onesixth}\begin{center}}{c}<{\end{center}\end{minipage}}}
\newcolumntype{E}{@{}l@{}} 
\def\checkmark{\tikz\fill[scale=0.4](0,.35) -- (.25,0) -- (1,.7) -- (.25,.15) -- cycle;}

\begin{document}

\title{Analysis and Design of 8-Bit CMOS Priority Encoders}

\numberofauthors{1} 
\author{
\alignauthor
Xiaoyu Wang and Yukang Feng\\
       \affaddr{University of Virginia}\\
       \email{\{xw5ce, yf4rs\}@virginia.edu}
}

\maketitle
\begin{abstract}
A comprehensive review and fair comparison of previous priority encoder (PE) designs over the past one and a half decades are presented using a 45 nm technology.  Further, potential limitations of existed PEs are identified, based on which we propose a robust PE design. The new PE is able to eliminate race condition and charge sharing problem which are suffered by almost all the previous designs. Besides, the proposed PE can also be used in comprising higher order PEs by incorporating a carefully designed look-ahead structure. Simulation results demonstrate that our design can achieve one of the best power and delay performance among previous PEs and are free from potential risks. 

\end{abstract}


\section{Introduction}

A Priority encoder (PE) is a basic but critical unit in digital systems, and it has been widely used in many applications, such as fixed and floating point units, comparators, incrementer/decrementer circuits, sequential address encoder of content addressable memories and so on. In a multi-bit PE, each bit is assigned a priority weighting according to its own position weighting. A logic-1 priority token is initially and temporarily given to all bits. While the bit with a higher priority accepts a logic-1 input, it will pass a logic-0 signal to update the priority tokens of those lower priority bits to disable their priority. Meanwhile, when any one bit accepts a logic-0 input, it will also lose its priority by definition. For each input pattern, only the bit keeping the logic-1 token can generate a logic-1 output, while all the other bits will get logic-0 outputs.

Since late 1990s, more than eight different PE designs were proposed to achieve lower power dissipation, shorter delay and also less complexity. However, critical issues like possible charge sharing and race condition which might lead to fatal breakdown were rarely addressed. Based on these observations, this paper first presents a fair comparison of different PEs using the updated 45 nm process technology in order to shed some light on how to choose proper PEs in different applications. Next, we identify major limitations for each of existing PE and verify our analysis through simulations. Furthermore, a robust PE architecture suitable for serial cascading is proposed to get rid of potential charge sharing and race problems and maintain a decent delay and power performance at the same time. 
\section{Previous Works and Limitations}
A series of PE designed were proposed during the past decade. In this section, a brief review of the history of PE's advancement is presented and some limitations neglected before are identified. 
\subsection{Related Works}

Delgado-Frias and Nyathi \cite{firstPE1998} designed a priority encoder that permits sequential passage of priority token from the highest priority primary input to the lowest priority input -- the disadvantage of this design being that the sequential passage of priority token encounters a delay of $\mathcal{O}(n)$, where 'n' represents the total number of primary inputs or outputs. To alleviate the linear increase in delay, Wang and Huang \cite{wang2000high} put forward two 8-bit priority encoder designs, comprising two 4-bit encoder blocks with the provision of an internal lookahead signal -- one of the designs extensively utilizes pMOS transistors while the other design widely deploys nMOS transistors. 

Kun \emph{et al}. \cite{kun2004power} came up with the design idea of an 8-bit priority encoder module, eliminating the need for sub-modules and internal lookahead signaling. While Huang \emph{et al}. \cite{huang2002design} proposed a serial cascading architecture to realize higher order priority encoders, with the lookahead output of a 8-bit encoder module serving as the lookahead input for the succeeding encoder block, Kun \emph{et al}. \cite{kun2004power} proposed a parallel priority-based cascading topology to implement larger size priority encoders. Mohanraj \emph{et al}. \cite{mohanraj2012power} presented a new 8-bit priority encoder design, which is in fact a refinement of Kun \emph{et al}.'s encoder design by exploiting shared logic to reduce the number of devices needed for physical realization. 

Huang and Chang \cite{huang2010full} introduced a new NOR- based priority encoder, where during the precharge phase of the clock, all the outputs are driven to logic-high state, and in the evaluation phase, based upon input request(s), the input that assumes a higher priority is enabled and its corresponding output is retained as logic high, while the other primary outputs are pulled to logic low. In this aspect, the Huang and Chang's design is similar to Huang \emph{et al}.'s pMOS-based priority encoder design. Panchal  \emph{et al}. \cite{panchal2013design} modified Huang and Chang's work and came up with a similar PE based on active-low logic, which implies that input(s) have to be logic low so as to activate the PE to produce a desired logic high output. 
\subsection{Limitations of Previous Works}

The above described PE designs were proposed in a time interval of more than one and a half decades, and thus implemented using different process technologies, including 90 nm, 250 nm and 900 nm, most of which are outdated. Therefore, the studied metrics might not be representative enough to reflect their performance disparity. To the best knowledge of the authors, there has not been any thorough and fair comparison of all the major PE designs in the past decade, i.e., utilizing the same updated process technology. For this consideration, we implement these PEs and measure main design metrics using the same 45 nm process technology, aiming to provide some insights for choosing different PEs under different application circumstances. 

Based on extensive and in-depth study of previous works, it is noticed that almost all the designs focused on improving the three metrics, i.e., power dissipation, worst-case delay and number of transistors, which are vital aspects for PE design. However, what they failed to consider is the robustness of these PEs, i.e., whether they could still function correctly in some extreme or untypical scenarios. For instance, with some certain input combinations, PEs designed without considering the possibility of charge sharing problem will generate outputs with lower voltage than logic high or even result in flipping. Another example is that due to unexpected delay of the look-ahead signal, a race condition might happen causing the stage losing priority still outputs logic high signal(s). These two cases will both severely compromise PE's robustness and thus limit their applicability or require more complexity when designing other parts of a system.  In addition, since a typical PE only consists of 8-bit inputs and higher order PEs are widely used in various systems, the ability of 8-bit PEs to comprise higher order ones though serial cascading is desired in most cases. However, some of the previous designs are not suitable for realizing higher order PEs, even though they included a look-ahead signal in their circuits. 

Table 1 summarizes major limitations of previous PE designs in terms of race condition, charge sharing problem and feasibility of cascade, in which checkmarks represent that corresponding designs suffer from those limitations.  From Table 1 we notice that all the previous PE designs failed to fix at least one of the problems except the high-speed PE proposed in \cite{wang2000high}, which on the hand, has another disadvantage that its power dissipation is much higher than the rest of PEs. Given these observations, we propose a robust PE architecture which is race-condition and charge-sharing free, as well as suitable for realizing higher order PEs.
\begin{table}
    \centering
    \caption{Limitations of different PEs}
    \begin{tabular}{|c|c|c|c|}
    	\hline
	\multicolumn{1}{|C|}{Limitations} & \multicolumn{1}{C|}{Race condition} & \multicolumn{1}{C|}{Charge sharing} & \multicolumn{1}{C|}{\tabincell{c}{Unsuitable \\ for cascade}} \\ 
        \hline
        \multicolumn{1}{|C|}{\tabincell{c}{Wang and \\ Huang 1 \cite{wang2000high} }} & \multicolumn{1}{C|}{ } & \multicolumn{1}{C|}{ } & \multicolumn{1}{C|}{ } \\ 
        \hline
        \multicolumn{1}{|C|}{\tabincell{c}{Wang and \\ Huang 2 \cite{wang2000high} }} & \multicolumn{1}{C|}{ } & \multicolumn{1}{C|}{\checkmark} & \multicolumn{1}{C|}{ } \\ 
        \hline
        \multicolumn{1}{|C|}{\tabincell{c}{Kun \\ \emph{et al}. \cite{kun2004power} }} & \multicolumn{1}{C|}{\checkmark} & \multicolumn{1}{C|}{\checkmark} & \multicolumn{1}{C|}{ } \\ 
        \hline
        \multicolumn{1}{|C|}{\tabincell{c}{Huang \& \\Chang \cite{huang2010full} }} & \multicolumn{1}{C|}{\checkmark} & \multicolumn{1}{C|}{\checkmark (flipping)} & \multicolumn{1}{C|}{\checkmark} \\ 
        \hline
        \multicolumn{1}{|C|}{\tabincell{c}{Mohanraj \\ \emph{et al}. \cite{mohanraj2012power} }} & \multicolumn{1}{C|}{\checkmark} & \multicolumn{1}{C|}{\checkmark} & \multicolumn{1}{C|}{ } \\ 
        \hline
        \multicolumn{1}{|C|}{\tabincell{c}{Panchal \\ \emph{et al}. \cite{panchal2013design} }} & \multicolumn{1}{C|}{\checkmark} & \multicolumn{1}{C|}{\checkmark (flipping)} & \multicolumn{1}{C|}{\checkmark} \\ 
        \hline
        \end{tabular}
\end{table}

\section{Proposed Robust PE}

In a multi-bit PE, the output of the $i$-th bit is $OP_i=IP_i\bm\cdot P_i$, where $IP_i$ is the corresponding input data and $P_i$ stands for the priority token passed onto this bit. When the input of the lower significant bit is 0, the priority token is passed onto the next bit, i.e., $P_i= \overline{IP}_{i-1}\bm\cdot P_{i-1}$. The general expression of outputs $OP_i$ can be written as
\begin{equation}
	OP_i = IP_i\bm\cdot \overline{IP}_{i-1}\bm\cdot \overline{IP}_{i-1}\bm\cdot \overline{IP}_{i-3}\bm\cdot \bm\cdot \bm\cdot \overline{IP}_{1}\bm\cdot \overline{IP}_{0}
\end{equation}
For the proposed 8-bit PE with a three-level look-ahead structure shown in Figure 1, the fundamental equations governing the PE are given as follows
\begin{align}
\begin{split}
 	OP_0 &= \overline{LA}\bm\cdot IP_0 \\
	OP_1 &= \overline{LA}\bm\cdot \overline{IP}_0\bm\cdot IP_1 \\
	OP_2 &= \overline{LA}\bm\cdot \overline{IP}_0\bm\cdot \overline{IP}_1\bm\cdot IP_2 \\
	OP_3 &= \overline{LA}\bm\cdot \overline{IP}_0\bm\cdot \overline{IP}_1\bm\cdot \overline{IP}_2\bm\cdot IP_3 \\
	LA_{inter} &= \overline{LA} + IP_0 + IP_1 + IP_2 + IP_3 \\
	OP_4 &= \overline{LA}_{inter}\bm\cdot IP_4 \\
	OP_5 &= \overline{LA}_{inter}\bm\cdot \overline{IP}_4\bm\cdot IP_5 \\
	OP_6 &= \overline{LA}_{inter}\bm\cdot \overline{IP}_4\bm\cdot \overline{IP}_5\bm\cdot IP_6 \\
	OP_7 &= \overline{LA}_{inter}\bm\cdot \overline{IP}_4\bm\cdot \overline{IP}_5\bm\cdot \overline{IP}_6\bm\cdot IP_7 
\end{split}
\end{align} 

When Clock becomes 0, the circuit is in the pre-discharge phase.  $LA_{inter}$ is 0 and all outputs are pre-discharged to 0. When Clock becomes 1, the circuit enters the evaluation phase. In the circuitry, the p-type dynamic gates for $OP_0\sim OP_3$ realize the first-level look-ahead functions with $la_0\sim la_2$ acting as the look-ahead signals. Owing to the first-level look-ahead structure, the four outputs $OP_0\sim OP_3$ evaluate at the same time.

$LA_{inter}$ is used to realize the second-level look-ahead function between the higher-priority and lower-priority 4-bit cells and \emph{LA} is used to realized the third-level look-ahead function to decide whether the current 8-bit macro cell owns the priority.  Note that the new design uses active-low look-ahead signals, which means that another stage with higher weighting owns the priority when \emph{LA} is logic 1. In such a case, $OP_0\sim OP_7$ will be set to logic 0 during the evaluation phase. If \emph{LA} is logic 0 to pass the priority into the current macro cell, $OP_0\sim OP_3$ are decided by $IP_0\sim IP_3$ directly, while $OP_4\sim OP_7$ are decided by both $IP_4\sim IP_7$ and the second-level look-ahead signal $LA_{inter}$.

There are a number of advantages of the new 8-bit PE cell over the conventional ones. First, the PE cell is designed to be race-free by using $rs0\sim rs7$. At the beginning of the evaluation phase, each output bit is evaluated immediately according to the input signals and at most one output bit will be charged from 0 to 1. However, these outputs may be incorrect. When the correct look-ahead signal arrives a little bit later than the rising edge of the clock signal, if the current stage owns the priority both signals \emph{LA} and $LA_{inter}$ will remain at 0 and the previously evaluated outputs are exactly correct. Otherwise, if the current stage loses it priority,  both signals \emph{LA} and $LA_{inter}$ will be 1 to turn on $rs0\sim rs7$ to enforce all the outputs of the current stage to be logic low. Second, because the circuit utilizes the three-level look-ahead-signal structure, it has the high-speed characteristics. Third, the PE design will not suffer from charge sharing, since there only exist two parallel NMOS transistors between each output and ground. Fourth, due to the series-type circuit structure, all outputs will evaluate in the evaluation phase but with only one output being charged after the pre-discharge phase and also only the output with high voltage will be discharged in the next pre-discharge phase. This means a significant reduction of the switching activity and the corresponding switching power. Last but not least, given the carefully designed look-ahead signal, the new PE could also be used as a macro cell for comprising higher order PEs by utilizing the parallel priority look-ahead architecture of Kun \emph{et al}. \cite{kun2004power}.
\begin{figure}\label{RobustPE}
\centering
	\includegraphics[width=3.4in,height=4.5in]{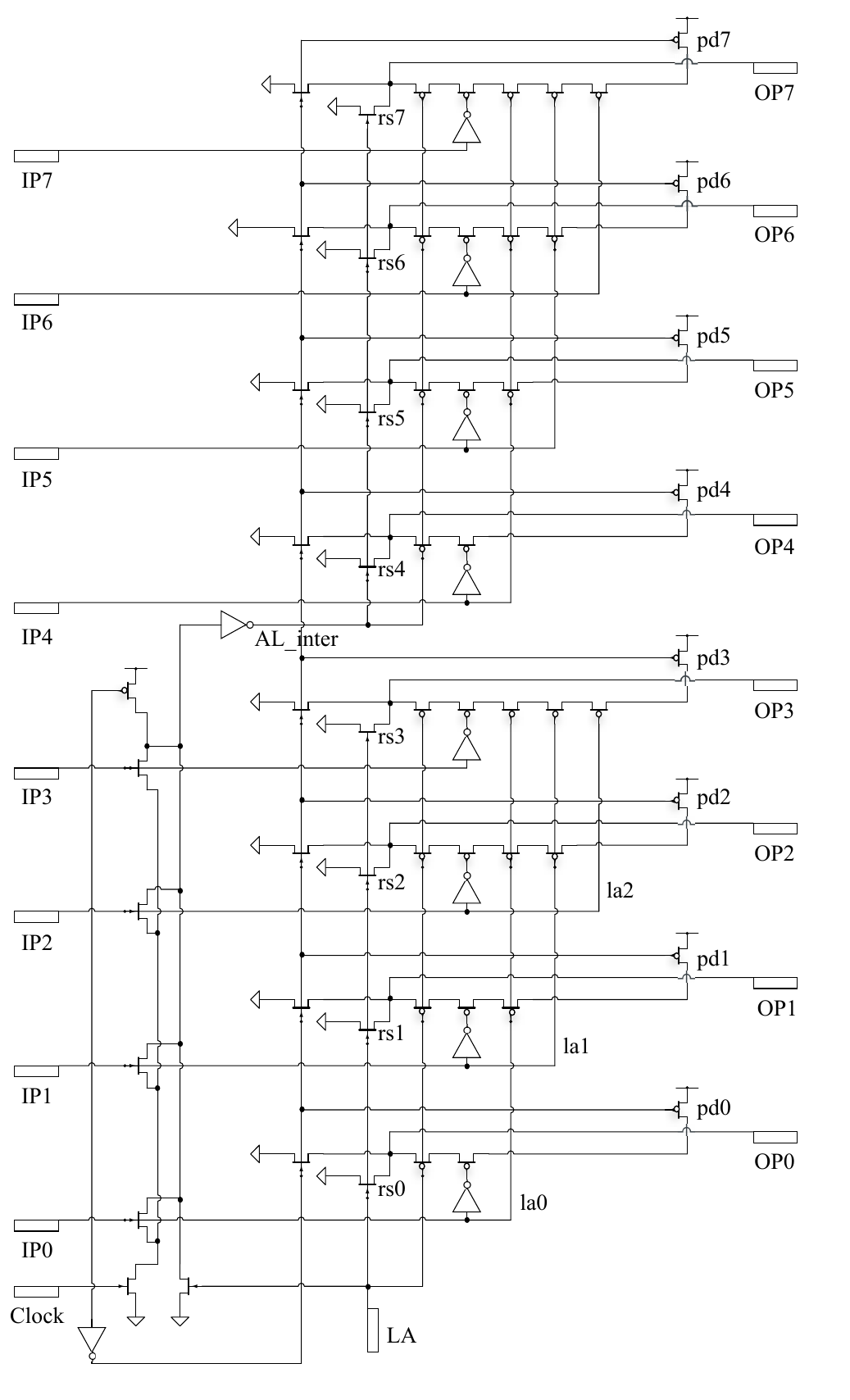}
\caption{Proposed 8-bit robust priority encoder.}
\end{figure}
\section{Performance Evaluation and Experimental Results}

\subsection{Fair Performance Comparison of PEs}
\begin{table*}
    \centering
    \caption{Comparison of Design parameters of different 8-bit dynamic CMOS PEs}
    \begin{tabular}{|c|c|c|c|c|c|c|c|}
    	\hline
	\multicolumn{1}{|C|}{\tabincell{c}{Design \\ metrics}} & \multicolumn{1}{C|}{ \tabincell{c}{Wang and \\ Huang 1 \cite{wang2000high} }} & \multicolumn{1}{C|}{ \tabincell{c}{Wang and \\ Huang 2 \cite{wang2000high} }} & \multicolumn{1}{C|}{ \tabincell{c}{Kun \\ \emph{et al}. \cite{kun2004power} }} & \multicolumn{1}{C|}{ \tabincell{c}{Huang \& \\Chang \cite{huang2010full} }} & \multicolumn{1}{C|}{ \tabincell{c}{Mohanraj \\ \emph{et al}. \cite{mohanraj2012power} }} & \multicolumn{1}{C|}{ \tabincell{c}{Panchal \\ \emph{et al}. \cite{panchal2013design} }} & \multicolumn{1}{C|}{ \tabincell{c}{New design}} \\ 
        \hline
	\multicolumn{1}{|C|}{\tabincell{c}{Power ($\mu$W)}} & \multicolumn{1}{C|}{79.11} & \multicolumn{1}{C|}{6.119}& \multicolumn{1}{C|}{9.422}& \multicolumn{1}{C|}{6.544}& \multicolumn{1}{C|}{3.189}& \multicolumn{1}{C|}{7.100}& \multicolumn{1}{C|}{6.879} \\
        \hline
        \multicolumn{1}{|C|}{\tabincell{c}{Delay (ns)}} & \multicolumn{1}{C|}{0.177} & \multicolumn{1}{C|}{0.346}& \multicolumn{1}{C|}{0.292}& \multicolumn{1}{C|}{0.281}& \multicolumn{1}{C|}{0.274}& \multicolumn{1}{C|}{1.018}& \multicolumn{1}{C|}{0.278} \\ 
        \hline
        \multicolumn{1}{|C|}{\tabincell{c}{PDP (fJ)}} & \multicolumn{1}{C|}{14.002} & \multicolumn{1}{C|}{2.117}& \multicolumn{1}{C|}{2.751}& \multicolumn{1}{C|}{1.839}& \multicolumn{1}{C|}{0.874}& \multicolumn{1}{C|}{7.228}& \multicolumn{1}{C|}{1.912} \\ 
        \hline
        \multicolumn{1}{|C|}{\tabincell{c}{Number of\\ transistors}} & \multicolumn{1}{C|}{102} & \multicolumn{1}{C|}{103}& \multicolumn{1}{C|}{62}& \multicolumn{1}{C|}{76}& \multicolumn{1}{C|}{55}& \multicolumn{1}{C|}{60}& \multicolumn{1}{C|}{79} \\ 
        \hline
        \end{tabular}
\end{table*}

Seven 8-bit dynamic CMOS PEs including the proposed design have been implemented at the transistor level and simulated using Cadence based on a 45nm CMOS process design kit from NCSU, i.e., FreePDK, with a supply voltage of 1.1V. A combination of all the possible inputs are applied at a clock frequency of 50 MHz to verify the functionality of these different PEs, as well as to estimate the average power dissipation. The total average power dissipation and critical path delay metrics of different 8-bit PEs are given in Table 2, along with the device count required for physical design. The device count, in terms of number of transistors needed, is assumed to be representative of the area occupancy of the circuit. 

From Table 2 we notice that the PE presented by Wang and Huang 1 \cite{wang2000high} is the fastest design, while its power dissipation, PDP and transistor number are much larger than the rest of the designs, which makes its much less desirable in applications. In terms of the four metrics considered here, the power, delay and area optimized PE proposed in \cite{mohanraj2012power} achieves the optimal overall performance, while it comes with the cost of possible race condition and charge sharing as discussed in Section II. The proposed robust PE has a balanced performance in these four aspects, i.e., with a relatively small number of transistors, the new PE has one of the smallest power consumption, delay and PDP. 
\subsection{Potential Failure of Previous PE Designs}

Given the identified potential limitations of previous PE designs, in the section, we provide some simulation results to confirm our analysis in Section II. For a PE design failing to consider potential race condition, i.e., the look-ahead signal that disables the current stage arrives after the clock edge starting a evaluation phase, the outputs might not be disabled immediately, leading to unexpected results. Here the power-optimized PE \cite{kun2004power} is used as an example and corresponding timing diagram is given in Figure 2. For this design with active-high look-ahead signal, when the look-ahead signal is logic 0, all the outputs of the current stage should be disabled no matter what the input values are. However, as shown in Figure 2, when the rising edge of Clock arrives, \emph{LA} is high and the second bit owning the priority outputs logic 1, while when \emph{LA} arrives later, $OP_1$ would not be disabled, leading to the possibility that more than one of the outputs have logic 1 in a higher order PE.
\begin{figure}\label{PowerOptimized_Race}
\centering
	\includegraphics[width=3.2in,height=1.8in]{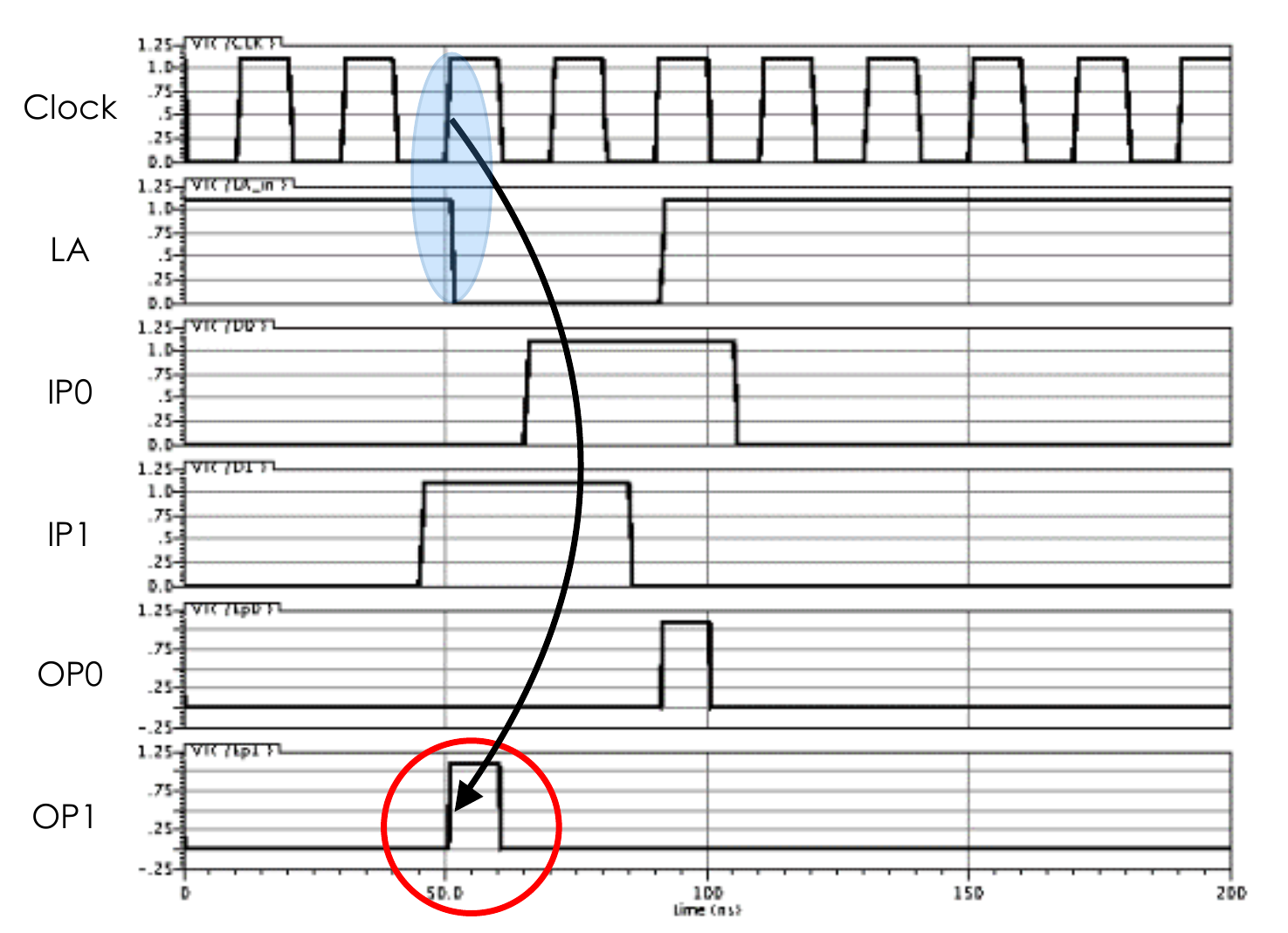}
\caption{Race condition in the power-optimized PE.}
\end{figure}

Next, we consider another fatal problem existing in previous works -- charge sharing. In Figure 3, a possible output flipping of the power, delay and area optimized PE proposed in \cite{mohanraj2012power} is presented. Under normal situations, if the input(s) with higher priority is logic 1, all the outputs of lower-priority bits should be logic 0, which is shown as in the green circle in Figure 3. However, given some special combinations of inputs, some outputs will be flipped, leading to severe malfunction of PEs. The output flipping due to charge sharing is displayed with the red circle in Figure 3, where the logic 1 of $OP_7$ is unexpected. 
\begin{figure}\label{PDA_ChargeSharing}
\centering
	\includegraphics[width=3.2in,height=1.8in]{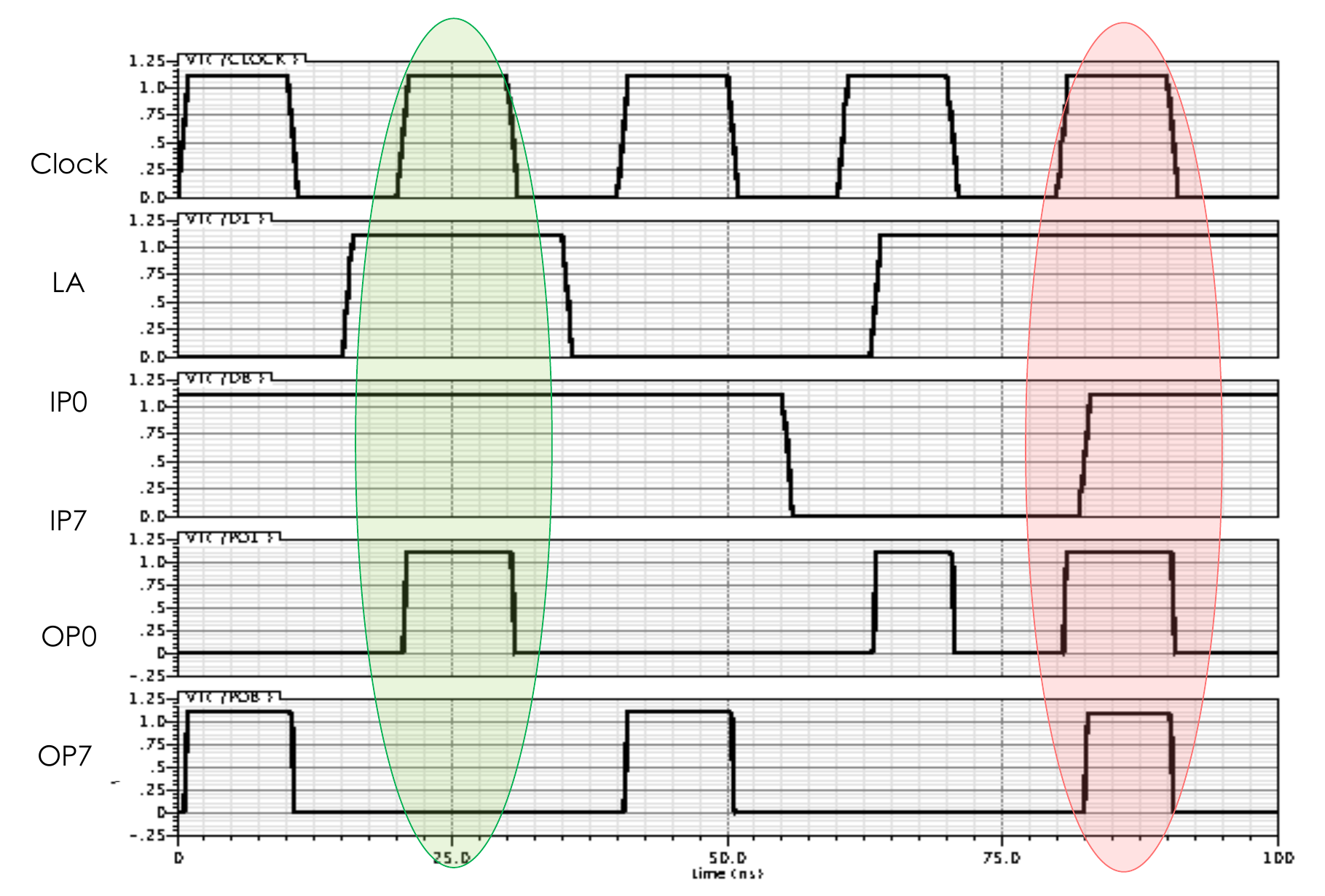}
\caption{Output flipping due to charge sharing in the power, delay and area optimized PE.}
\end{figure}
\subsection{Robustness of the Proposed PE}

The advancement of the proposed PE is mainly reflected in three aspects, i.e., free of race condition, charge sharing and the suitability for comprising higher order PEs. The latter two advantages could be verified straightforward given the fact that the outputs are connected to the lowest part of the circuit and that the third-level look-ahead signal $LA$ is adopted to realize cascading in higher order PEs. Here we only present the simulation result in Figure 4 to validate the race-condition-free property of the new PE. 

Consider the scenario given in the left green circle in Figure 4. At the beginning of the evaluation phase, the current stage owns the priority ($LA$ is logic 0), $OP_3$ is charged. Then $LA$ arrives later, which will discharge $OP_3$ immediately, outputting the correct results. Again, when the current stage changes from disabled to enabled due to the arrival of a logic 0 $LA$, $OP_1$ corresponding to the bit which owns priority will immediately turned to logic 1 without waiting for the next evaluation phase. Due to careful design of the look-ahead structure, any potential erroneous arrival time of the look-ahead signal due to problematic timing design of other parts of a system will be fixed within the PE, without passing fault outputs to following levels.
\begin{figure}\label{NewDesign_RaceFree}
\centering
	\includegraphics[width=3.2in,height=1.8in]{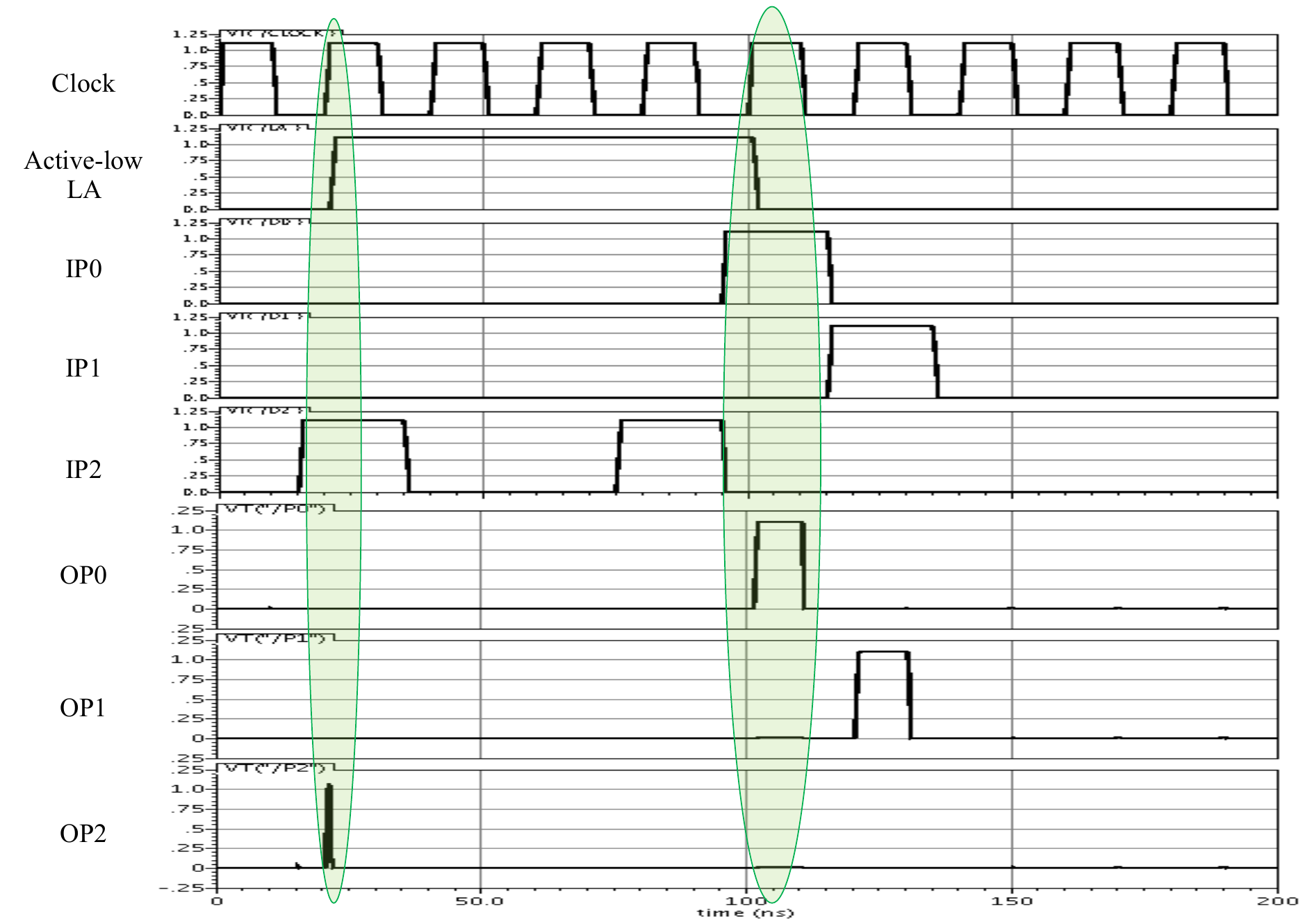}
\caption{Robustness of the proposed PE against race condition.}
\end{figure}

\section{Conclusions}

A comprehensive review of existing PEs and their fair performance comparison are presented using a 45 nm technology in terms of three design metrics -- power, delay and number of transistors. Moreover, we analyze these designs closely to identify their potential limitations, including possible race and charge sharing problems, and infeasibility to comprising higher order PEs. Our analysis shows that almost all the existing PEs suffer from one or more of these disadvantages. In order to obtain a PE which is capable of overcoming these shortages, a robust PE structure is proposed, which is validated to be charge-sharing and race-condition free, proper for cascading and also have one of the best power and delay performance.
%
\bibliographystyle{abbrv}
\bibliography{Reference}  
%
%

\end{document}